\documentclass[utf8]{FrontiersinHarvard} 

\usepackage{url,hyperref,lineno,microtype,subcaption}
\usepackage[onehalfspacing]{setspace}
\usepackage{lipsum}  
\usepackage{aas_macros}
\usepackage{CJKutf8}
\newcommand{\cntext}[1]{\begin{CJK}{UTF8}{gbsn}#1\end{CJK}\kern-1ex}

\def\keyFont{\fontsize{8}{11}\helveticabold }
\def\firstAuthorLast{L\"{o}rin\v{c}\'{i}k {et~al.}}
\def\Authors{Juraj L\"{o}rin\v{c}\'{i}k\,$^{1,2,*}$, Vanessa Polito\,$^{1,2,3}$, Bart De Pontieu\,$^{2,4,5}$, Sijie Yu (\cntext{余思捷}~~)\,$^{6}$, Nabil Freij\,$^{1,2}$}
\def\Address{$^{1}$Bay Area Environmental Research Institute, NASA Research Park, Moffett Field, CA 94035, USA \\
$^{2}$Lockheed Martin Solar \& Astrophysics Laboratory, Org. A021S, Bldg. 252, 3251 Hanover St., Palo Alto, CA 94304, USA\\
$^{3}$Department of Physics, Oregon State University, 301 Weniger Hall, Corvallis, OR 97331\\
$^{4}$Institute of Theoretical Astrophysics, University of Oslo, P.O. Box 1029 Blindern, 0315 Oslo, Norway\\
$^{5}$Rosseland Centre for Solar Physics, University of Oslo, P.O. Box 1029 Blindern, 0315 Oslo, Norway\\
$^{6}$Center for Solar-Terrestrial Research, New Jersey Institute of Technology, 3L King Jr. Blvd., Newark, NJ 07102-1982, USA}

\def\corrAuthor{J. L\"{o}rin\v{c}\'{i}k}
\def\corrEmail{lorincik@baeri.org}

\begin{document}
\onecolumn
\firstpage{1}

\title[]{Rapid variations of Si IV spectra in a flare\\ observed by IRIS at a sub-second cadence} 

\author[\firstAuthorLast ]{\Authors} 
\address{} 
\correspondance{} 

\extraAuth{}

\maketitle

\begin{abstract}

We report on observations of highly-varying Si IV 1402.77\,\AA~line profiles observed with the Interface Region Imaging Spectrograph (IRIS) during the M-class flare from 2022 January 18 at an unprecedented 0.8\,s cadence. Moment analysis of this line observed in flare ribbon kernels showed that the intensity, Doppler velocity, and non-thermal broadening exhibited variations with periods below 10\,s. These variations were found to be correlated with properties of the Gaussian fit to a well-resolved secondary component of the line redshifted by up to 70\,km\,s$^{-1}$, while the primary component was consistently observed near the rest wavelength of the line. A particularly high correlation was found between the non-thermal broadening of the line resulting from the moment analysis and the redshift of the secondary component. This means that the oscillatory enhancements in the line broadening were due to plasma flows (away from the observer) with varying properties. A simple de-projection of the Doppler velocities of the secondary component based on a three-dimensional reconstruction of flare loops rooted in the kernel suggests that the observed flows were caused by downflows and compatible with strong condensation flows recently predicted by numerical simulations. Furthermore, peaks of the intensity and the trends of Doppler velocity of the Gaussian fit to the secondary component (averaged in the ribbon) were found to correspond to one of the quasi-periodic pulsations (QPPs) detected during the event in the soft X-ray flux (as measured by the Geostationary Operational Environmental Satellite, GOES) and the microwave radio flux (as measured by the Expanded Owens Valley Solar Array, EOVSA). This result supports a scenario in which the QPPs were driven by repeated magnetic reconnection. 

\tiny
 \keyFont{ \section{Keywords:} Solar flares (1496), Solar atmosphere (1477), Solar ultraviolet emission (1533), Solar transition region (1532), Solar magnetic reconnection (1504)} 
\end{abstract}

\twocolumn

\section{Introduction} \label{sec_intro}

Spectroscopic observations of solar flare ribbons provide a wealth of information about the transfer of energy between the site of energy release in the corona and the lower atmosphere of the Sun \citep[see e.g. the review of][]{fletcher11}. A typical manifestation of this process are Doppler shifts of spectral lines which formation region is affected by upflows and downflows of plasma \citep[see e.g.][]{doschek80, canfield87, brosius03} along magnetic field lines forming as a consequence of magnetic reconnection, the driver of solar flares \citep[see Section 1 in][]{shibata11}. The upflows, which can be identified as blueshifts of lines forming at coronal and flare temperatures (log \mbox{($T$ [K])  $\approx 6 - 7$}), are due to the chromospheric evaporation \citep[e.g.][]{acton82, delzanna06, brosius13}. In this process, the deposited energy heats plasma at chromospheric heights which consequently expands at speeds sometimes reaching hundreds km\,s$^{-1}$ \citep[see e.g. the review of][]{milligan15}, and fills flare loops. The downflows, observable as redshifts of lines formed at lower temperatures in the chromosphere and the transition region (log \mbox{($T$ [K]) $\approx 4-5$}), are due to the process termed chromospheric condensation \citep[e.g.][]{fischer85a, warren16, graham20}. Chromospheric condensation occurs below the expanding region and the speeds of the downflows driven by the expansion are typically an order of magnitude lower than those of the evaporation, reaching a few to few tens km\,s$^{-1}$. Therefore, while the evaporation can in some cases lead to complete blueshifts of hot lines \citep[e.g.][]{young15, dudik16, polito16a}, the condensation-induced downflows are often only observed as modest redshifts or enhancements in red wings of cooler lines \citep{tian15, warren16, brosius18, yu20, lorincik22}. The Doppler shifts indicative of the evaporation and condensation evolve in time. Apart from a slight delay between the visibility of the earliest traces of the condensation and the evaporation, the absolute values of the corresponding red and blue Doppler velocities drop over tens and hundreds of seconds, respectively \citep[][]{grahamcauzzi15, graham20}. 

Another widely-discussed observable which traces the energization of the lower solar atmosphere is the broadening of line profiles. Apart from the thermal and instrumental broadening, line profiles are often subject to excess non-thermal broadening \citep[e.g.][]{dere93, harra13, stores21}. In flares, the non-thermal broadening is usually attributed to unresolved turbulent motions in the emitting region or a superposition of multiple sources of Doppler-shifted emission along the line-of-sight \citep[e.g.][]{milligan11, doschek14, warren18}. Non-thermal broadening and its origins have been studied in observations of the Interface Region Imaging Spectrometer \citep[IRIS;][]{depontieu14} sometimes supplemented with flare simulations performed using the RADYN code \citep[][]{carlsson92, carlsson95, carlsson97, allred15}. The high spectral, spatial, and temporal resolutions of IRIS are well suited to study highly-varying flare spectra \citep[for a summary of results of IRIS see the review of][]{depontieu21}. For example, \citet{polito19} showed that the superposition of flows alone cannot account for broad profiles of the Fe XXI 1354.1\,\AA~line observed in flare ribbon. \citet{jeffrey18} discovered an increase and oscillations of non-thermal broadening of the Si IV 1402.77\,\AA~line in a ribbon which formed during a small B-class flare. Since these occurred prior to the impulsive phase of the flare determined from the trend of the X-ray emission, the authors conclude that the turbulence contributed to plasma heating. This is supported by the observed $\approx$10\,s period of the oscillations which corresponds to the timescale of the dissipation of turbulent energy predicted to range between 1 and 10\,s \citep{kontar17}. Oscillations of both excess width and intensity of the same line of Si IV with periods between 5 and 10\,s were also reported in the analysis of two microflares of \citet{chitta20}. The authors proposed that the turbulence acts as a triggering mechanism of fast magnetic reconnection. The broadened profiles are also often subject to Doppler shifts \citep[e.g.][]{jeffrey18}, in some cases clearly attributed to the evaporation and condensation \citep[e.g.][]{milligan11, li15, polito15}. This indicates that these processes might act simultaneously. 

The results of these studies suggest that turbulence plays an important role during the deposition of the energy released during flares in the lower solar atmosphere, possibly contributing to the heating itself. To our knowledge, in IRIS observations the signatures of the turbulence as well as high-frequency ($P < 10$\,s) oscillations of properties of lines formed in the transition region have only been studied in datasets containing spectra observed during small flares (see e.g. the references above) and active region brightenings \citep{gupta18}. This is most-likely caused by the scarcity of major flare events observed by recent instrumentation at a cadence that would allow for the investigation of these phenomena. Thanks to the recent development of very high-cadence IRIS observing programs, new opportunities for advances in this topic are now possible as IRIS has routinely been observing transient solar activity at a sub-second cadence, starting in Fall 2021. In this work we detail high-frequency oscillations of properties of the Si IV 1402.77\,\AA~line, including its non-thermal broadening, observed during the M-class flare from 2022 January 18. 

This manuscript is structured as follows. In Section \ref{sec_event} we briefly describe the investigated event. Analysis of spectroscopic and imaging data is detailed in Sections \ref{sec_spectra} and \ref{sec_loops3D}. In Section \ref{sec_radio} we focus on relating the time evolution of spectra observed by IRIS to features visible in SXR and microwave radio lightcurves. A brief summary of our results is provided in Section \ref{sec_summary}.

\section{2022 January 18 flare} \label{sec_event}

\subsection{Data}

In this paper we focus on spectroscopic observations with IRIS of the 2022 January 18 M1.5-class flare. The IRIS instrument observes solar spectra in two far ultraviolet (FUV) bands at 1331.6\,\AA~-- 1358.4\,\AA~and 1380.6\,\AA~-- 1406.8\,\AA~and a near ultraviolet (NUV) band at 2782.6\,\AA~-- 2833.9\,\AA. These bands contain a multitude of lines formed across the solar atmosphere at temperatures ranging between \mbox{log $T$ [K] = 3.7 and 7}. The spatial and spectral resolutions in the FUV band, containing the Si IV 1402.77\,\AA~line that we analyze here, reach 0.33" and 0.026\,\AA, respectively. The pixel size of IRIS is 0.167". This flare was captured in the sit-and-stare mode with spectral and spatial summing of 2. This event is the first major (GOES-class) flare IRIS observed at an unprecedented sub-second cadence of 0.8\,s. The exposure time employed in this observation is 0.3\,s.

The spectrograph datacube is accompanied by a series of imaging observations from the IRIS Slit-Jaw Imager (SJI) at 2796\,\AA. The analyzed flare is further studied using imaging data from the Atmospheric Imaging Assembly \citep[AIA;][]{lemen12} on-board the Solar Dynamics Observatory \citep[SDO;][]{pesnell12} and the Extreme Ultraviolet Imager \citep[EUVI;][]{wuelser04, howard08} on-board STEREO-A. Apart from the spectroscopic data we also investigate SXR lightcurves obtained by the Geostationary Operational Environmental Satellite (GOES) and microwave emission measured by the Expanded Owens Valley Solar Array \citep[EOVSA;][]{gary18}. 

\subsection{Description of the event} \label{sec_flare}

\begin{figure*}[h!]
\begin{center}
\includegraphics[width=18cm]{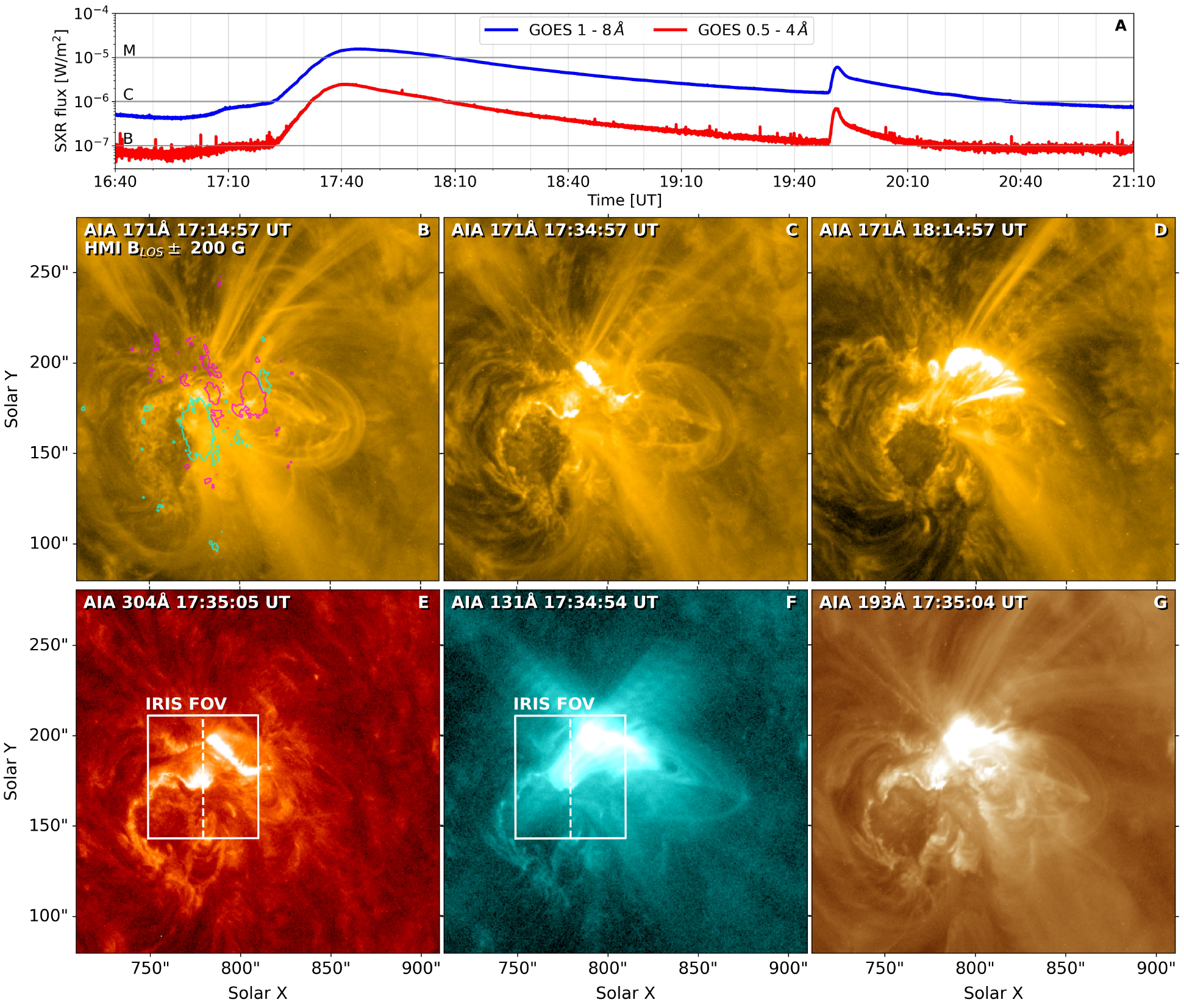}
\end{center}
\caption{Context observations of the 2022 January 18 flare. Panel (A) displays the evolution of the soft X-ray flux measured in the 0.5 -- 4\,\AA~(red) and 1 -- 8\,\AA~(blue) channels of the GOES satellite. Imaging observations of the flare in AIA 171\,\AA~during the onset, impulsive, and gradual phases of the flare is shown in panels (B) -- (D). The bottom row of the figure details the flare during its impulsive phase. Panel (E) shows flare ribbons observed in AIA 304\,\AA. Panels (F) and (G) detail hot flare emission as observed in AIA 131\,\AA~and AIA 193\,\AA, respectively. }\label{fig_overview}
\end{figure*}

Figure \ref{fig_overview} presents context observations of this event. Panel (A) depicts the soft X-ray flux measured in the two passbands of the GOES-16 satellite corresponding to 0.5 -- 4\,\AA~({red}) and 1 -- 8\,\AA~({blue}). According to this panel, the onset of the flare corresponded roughly to 17:00 UT, its impulsive phase lasted between $\approx$17:20 and $\approx$17:40 UT when the flux peaked. At this point the flux started to drop during the gradual phase lasting roughly until 20:00 UT when another small flare elsewhere on the disk occurred. 

The second row of this figure contains observations of the flare after its onset (panel (B)) as well as during its impulsive (panel (C)) and gradual (panel (D)) phases in the 171\,\AA~filter channel of AIA which is primarily sensitive to plasma emitting at log $T$ [K] = 5.8 \citep{lemen12}. The flare occurred in NOAA 12929 active region, which was located relatively close to the western limb at solar $x\approx$800". The magnetic field configuration of this active region is outlined using the green and magenta contours in panel (B), indicating the strength of the line-of-sight (LOS) component of the photospheric magnetic field as measured by the Helioseismic and Magnetic Imager \citep[HMI;][]{scherrer12} onboard SDO. The contours correspond to $\pm 200$\,G with the green contours corresponding to the positive polarity flux concentrations. During the impulsive phase of the flare (panel (C)), a pair of ribbons was observed to form, one to the south and one to the north. The comparison of panels (B) and (C) reveals that the southern ribbon was spatially coincident with the positive polarity flux concentrations while the northern one corresponded to the negative polarity flux. Here we further focus on spectra formed in the southern ribbon during the impulsive phase of the flare, also shown as observed in AIA 304\,\AA~(log $T$ [K] = 4.7), panel (E) when it crossed the slit of IRIS (white dashed line). Figure \ref{fig_overview}(F) shows an arcade of flare loops imaged in the AIA 131\,\AA~(log $T$ [K] = 7.0) which was forming at roughly the same time as shown in panel (E). The hot emission of the developing flare loop arcade is also shown in panel (G) where it is imaged in the AIA 193\,\AA. During flares, this channel is primarily sensitive to Fe XXIV emission peaking at log $T$ [K] = 7.25  \citep[see e.g.][]{odwyer10}.

\begin{figure*}[h!]
\begin{center}
\includegraphics[width=18cm]{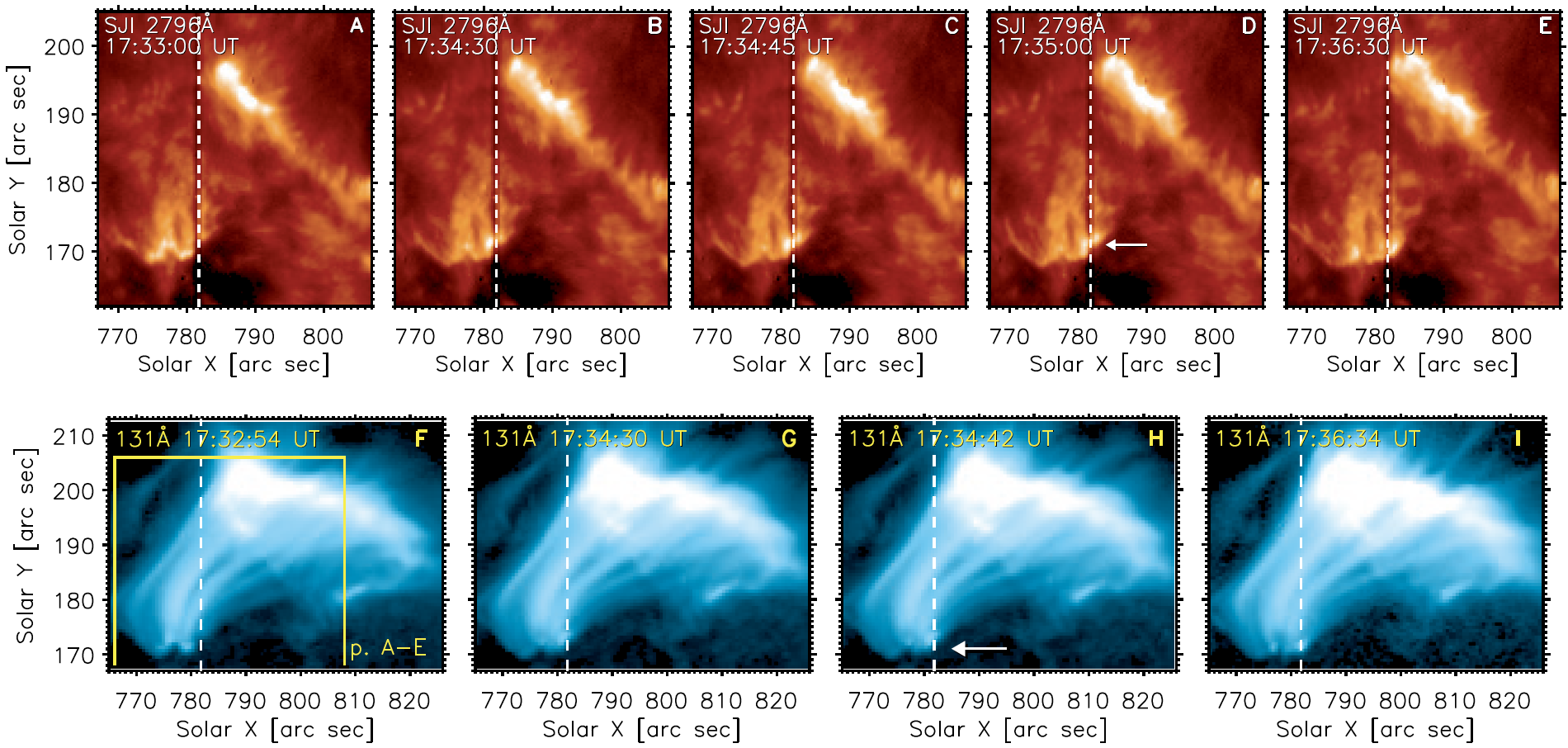}
\end{center}
\caption{A pair of flare ribbons (panels (A) -- (E)) and an arcade flare loops (panels (F) -- (I)) observed in the SJI 2796\,\AA~and AIA 131\,\AA~during the impulsive phase of the flare. White dashed line marks the position of the IRIS slit at 17:35 UT. The white arrow plotted in panels {(D)} and (H) points to kernels (top row), resp. flare loop footpoints (bottom row) which spectra are analyzed in this manuscript. \\Animated version of the SJI observations is available online. The animation covers the period between 17:30 and 17:40 UT.} \label{fig_loops_kernels}
\end{figure*}

A detailed view of the ribbons as well as the arcade of flare loops is provided in Figure \ref{fig_loops_kernels}. The structure of the ribbons is illustrated in the 2796\,\AA~SJI images during the time period that we focus on in this manuscript. The SJI data were averaged over 5 consecutive exposures to suppress the noise. An animation consisting of non-averaged full-cadence SJI observations of the ribbons is available online. The bright portion of the southern ribbon was composed of a series of small kernels. The time evolution of this ribbon, as depicted by the first row of Figure \ref{fig_loops_kernels} and the accompanying animation, was characterized by the motion of the kernels roughly directed from the solar east (panel (A)) towards the solar west direction (panel (E)), crossing the slit between $\approx$17:34 and $\approx$17:37 UT (panels (C), (D)). AIA 131\,\AA~observations of the flare arcade (panels (F) -- (I)) show several well-defined flare loops originating from these kernels. These flare loops are further discussed in Section \ref{sec_loops3D}. The white arrow plotted in panels (D) (and (H)) points toward the kernels (and flare loop footpoints) which, when passing through the slit, led to a major increase of the intensity, Doppler velocity, and broadening of the Si IV 1402.77\,\AA~line (Section \ref{sec_mom_maps}). We discuss this in the following Section.

\section{High-cadence IRIS observations of the Si IV 1402.77\,\AA~line} \label{sec_spectra}

In this manuscript we detail characteristics of the Si IV 1402.77\,\AA~line formed in the solar transition region. It was shown that{ especially in stronger flares the Si IV spectra can be affected by opacity effects \citep{kerr19, zhou22}. These are usually studied using the intensity ratio of the 1402.77\,\AA~line with the Si IV 1393.75\,\AA~line, predicted to be $\approx$2 when Si IV is formed under optically thin conditions \citep[see e.g.][]{mathioudakis99, gontikakis18}. As was recently reported by \citet{zhou22}, this ratio is also a function of wavelength, yielding different values across profiles of the two Si IV lines and large deviations in line centroids containing dips or reversals found in flare loops. Since we focus solely on ribbon spectra in which we did not observe typical spectral features associated with strong opacity effects \citep[see also][]{yan15, joshi21}, we assumed the line to be optically thin. The dataset we use also does not contain the Si IV 1393.75\,\AA~line for which we could not investigate the variations of the intensity ratio of the two Si IV lines, if any.}

\subsection{Moment analysis} 

\subsubsection{Maps of moments} \label{sec_mom_maps}

We begin the analysis of the Si IV 1402.77\,\AA~line by inspecting maps of the total line intensity (which is dependent on the peak intensity and the width of the profile), the Doppler velocity ($v_{\text{D}}$), and the non-thermal broadening ($v_\text{nt}$) as a function of time. These quantities were determined by calculating the moments of this line in the wavelength range of \mbox{$\lambda_0 \pm 0.6$\,\AA}~which contained the entire profile even in the pixels where the Si IV line was the broadest. Prior to the calculation of the moments, the FUV continuum averaged outside of this wavelength range was subtracted from the observed spectrum. {To obtain the Doppler velocities we used the reference wavelength of 1402.770\,\AA~\citep{sandlin86}.} Note that in this section as well as the remainder of the manuscript we only focused on pixels where the summed intensity exceeded 10\,DN. The non-thermal broadening was calculated using the formula \citep[see e.g.][]{testa16}: 
\begin{equation}
v_{\mathrm{nt}} = \sqrt{w_{\mathrm{1/e}}^2 - w_{\mathrm{inst}}^2  -  w_{\mathrm{th}}^2} .
\end{equation}
There, $w_{\text{1/e}}$ is the $1/e$ observed width of the line obtained as $\sqrt{2}\sigma$ of the fitting Gaussian, $w_\text{inst}$ stands for the instrumental broadening of 3.9\,km\,s$^{-1}$ \citep{depontieu14}, and $w_\text{th}$ is the thermal broadening which, for $T_\text{ion}$ = 80,000\,K, equals roughly 6.9\,km\,s$^{-1}$ \citep{depontieu15}.

These maps, shown in Figure \ref{fig_mom_maps}(A) -- (C), cover the time interval between 17:24 -- 17:44 UT when the Si IV emission in the southern ribbon was detectable. The map of the integrated intensity (panel (A)) shows a brightening along the ribbon persisting roughly between $\approx$17:34 and $\approx$17:37 UT, corresponding to the period when the bright kernels crossed the slit of IRIS (Section \ref{sec_flare}). This brightening has a relatively short onset, peaks at $\approx$17:35 UT, and then diminishes gradually over the next two minutes. The map of the Doppler velocity (panel (B)) indicates that, in the ribbon under  study, the line was usually redshifted with the highest Doppler velocities corresponding to $\approx$40\,km\,s$^{-1}$. The spatial and the temporal distribution of the Doppler velocities is less homogeneous than that of the intensities. The largest redshifts are found in two regions, one visible after $\approx$17:34 UT at $Y \approx 171$" and the other approximately three minutes later at $y \approx 169$". These two spatio-temporal regions are even more evident in the $v_\text{nt}$ map (panel (C)), suggesting that the line exhibited major non-thermal broadening in two separate episodes. 

The inclusions plotted in Figure \ref{fig_mom_maps}(A) -- (C) provide zoomed-in views of the maps during their initial increase after $\approx$17:34 UT {induced by the passage of the kernels through the slit detailed in Section \ref{sec_flare}. The grey and black arrows plotted therein indicate cuts used to investigate the time evolution of spectra at different pixels along the slit where the kernel emission was detected.}

\subsubsection{High-frequency variations of line moments} \label{sec_moments}

Figure \ref{fig_microscopic}(A) -- (C) shows variations of the summed intensity (grey), Doppler velocity ($v_{\text{D}}$, magenta), and the non-thermal broadening ($v_\text{nt}$, green) at three slit positions corresponding to the pixels indexed 83 -- 85 {(Solar $Y$ = 170.92 -- 171.59")} in the level 2 data cube. The time period shown corresponds to the initial increase of the intensity, and the first and second moment with a duration of roughly 80\,s. These three curves exhibit an overall increase followed by a decrease, best visible in the intensity trends at the pixel positions 84 and 85. While lasting {for only slightly more than a minute}, in what follows this trend will be referred to as the \textit{{long-term}} evolution (see Section \ref{sec_radio}). 

Apart from the {long-term} evolution, these plots contain a pattern consisting of brief enhancements (peaks) of the three properties of the line occurring at a relatively-high frequency. This pattern, a characteristic of the \textit{{short-term}} evolution, can be distinguished in all pixel positions plotted in Figure \ref{fig_microscopic}(A) -- (C). Unlike the {long-term} evolution, these enhancements are the most pronounced in the $v_{\text{D}}$ (magenta) and $v_\text{nt}$ (green) curves that exhibit very similar trends, with only some of them having a counterpart in the intensity curve (grey). The {heights} of these peaks are the largest at the pixel 84 plotted in panel (B) for the $v_{\text{D}}$ curve. {Magenta arrows were used to indicate some of the peaks along this curve in panel (B).} The period of the variations of $v_{\text{D}}$ and $v_\text{nt}$ estimated by counting the peaks between 17:34:30 and 17:35:30 UT is around 7\,s. The intensity exhibits fewer oscillations than the other two quantities (six as opposed to nine), leading to an estimated period of $\approx$10\,s.  

As stated in Section \ref{sec_intro}, quasi-periodic broadening during a flare of the Si IV 1402.77\,\AA~line, with a period of $\approx$10\,s, was first reported by \citet{jeffrey18}. A comparison of the upper panel of Figure 3 therein and our Figure \ref{fig_microscopic}(B) reveals that the trends exhibited by the non-thermal broadening are very similar. In both events the non-thermal broadening of the line had grown and exhibited oscillations before the peak of the intensity. In \citet{jeffrey18} these oscillations also preceded the increase of the {6 -- 12} keV X-ray flux measured by RHESSI, whereas the {long-term} evolution we report on (Section \ref{sec_radio}) was co-temporal with the increase of the SXR flux measured by GOES. The lifetime of the oscillations of $v_\text{nt}$ in the event analyzed here was also longer, as they persisted even after the peak of the line's intensity (see e.g. Figure \ref{fig_microscopic}(B)). Another important difference in the properties of the line between the two events is that the oscillations in broadening reported by \citet{jeffrey18} had no counterpart in the Doppler velocity and the intensity of the line. This finding motivated us to investigate whether the {short-term} evolution of the curves in Figure \ref{fig_microscopic}(A) -- (C) could simply be related to the motion of smaller kernels (not seen in Figure \ref{fig_loops_kernels}(A) -- (E)) through the slit. Lightcurves produced using full-cadence 2796\,\AA~SJI observations at and near the location where the resolved kernels entered the slit (Section \ref{sec_flare}) however did not exhibit enhancements relatable to the {short-term} evolution of properties of the Si IV line (not shown). This may be caused by the fact that the emission measured in the 2796\,\AA~SJI filter is dominated by the chromospheric Mg II k line formed below the formation region of Si IV. 

\subsection{Two-component profiles of the Si IV 1402.77\,\AA~line} 

Inspired by \citet{depontieu10}, we investigated whether the oscillations could be caused by quasi-periodic appearance of strongly Doppler-shifted components in the wings of the line, and their effect on the overall moments of the spectral line. We found that a portion of profiles observed in the analyzed ribbon exhibit two well-defined components. These profiles showed a larger spectral separation than that found in more typical flare observations of Si IV profiles consisting of a superposition of near-at-rest and Doppler-shifted components \citep[e.g.][]{li15, warren16, yu20, lorincik22}.

\begin{figure}[h!]
\includegraphics[width=85mm]{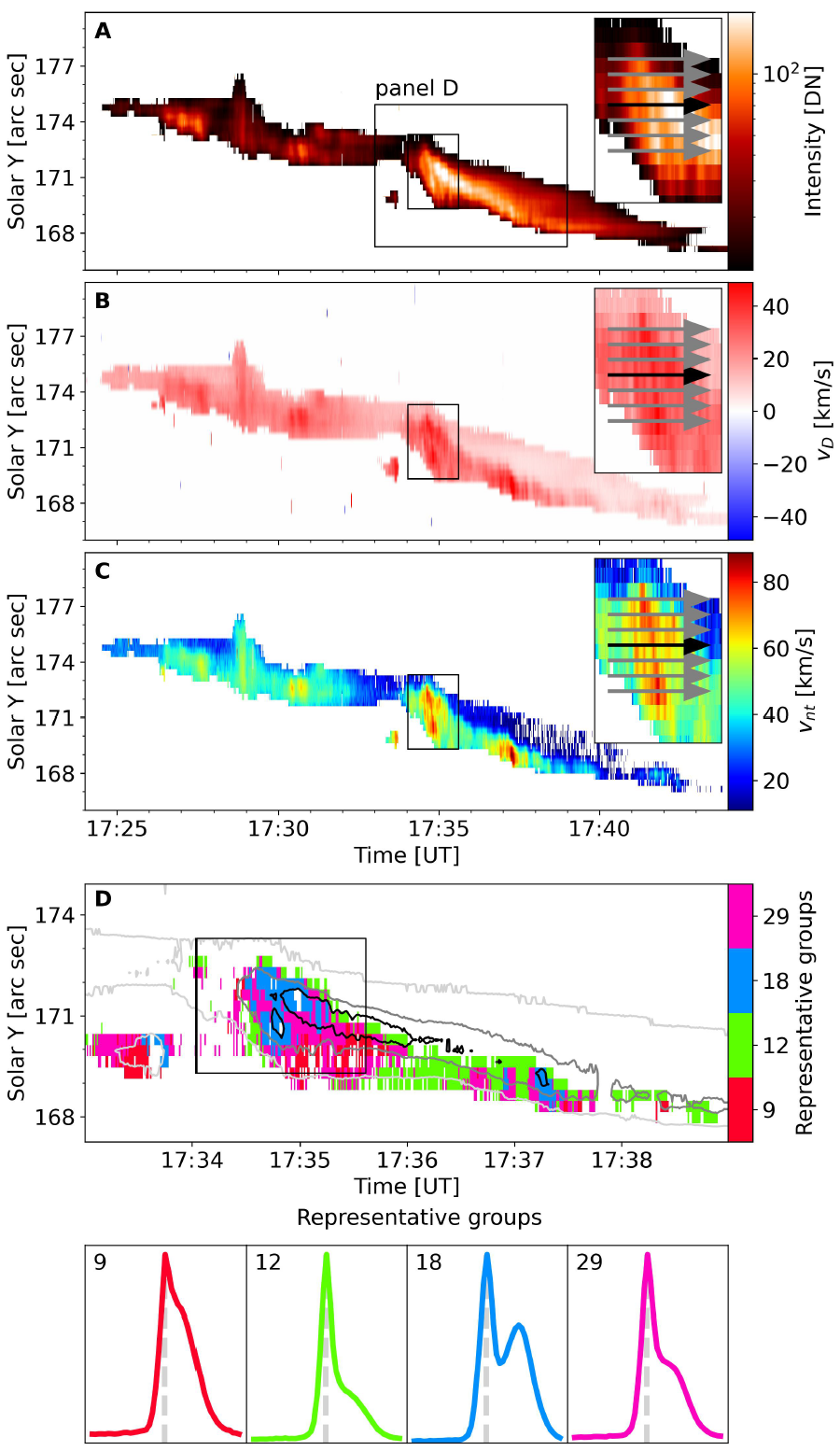}
\caption{Maps of the integrated intensity (panel (A)), Doppler velocity $v_{\text{D}}$ (panel (B)), and the non-thermal broadening $v_\text{nt}$ (panel (C)) of the Si IV 1402.77\,\AA~line resulting from the moment analysis. Arrows plotted in the {zoomed-in} inclusions indicate datapoints examined in Figures \ref{fig_microscopic} -- \ref{fig_correl}, {with the black arrow corresponding to the pixel position 84}. Panel (D) details the spatial distribution of Si IV line profiles which exhibited red wing enhancements. The individual pixels, color-coded according to the representative groups to which these profiles belong (displayed in the last row of the figure), are plotted atop of line intensity contours corresponding to 10 (light-grey), 70 (grey), and 140\,DN (black). }\label{fig_mom_maps}
\end{figure}

\begin{figure*}[h!]
\centering
\includegraphics[width=18cm]{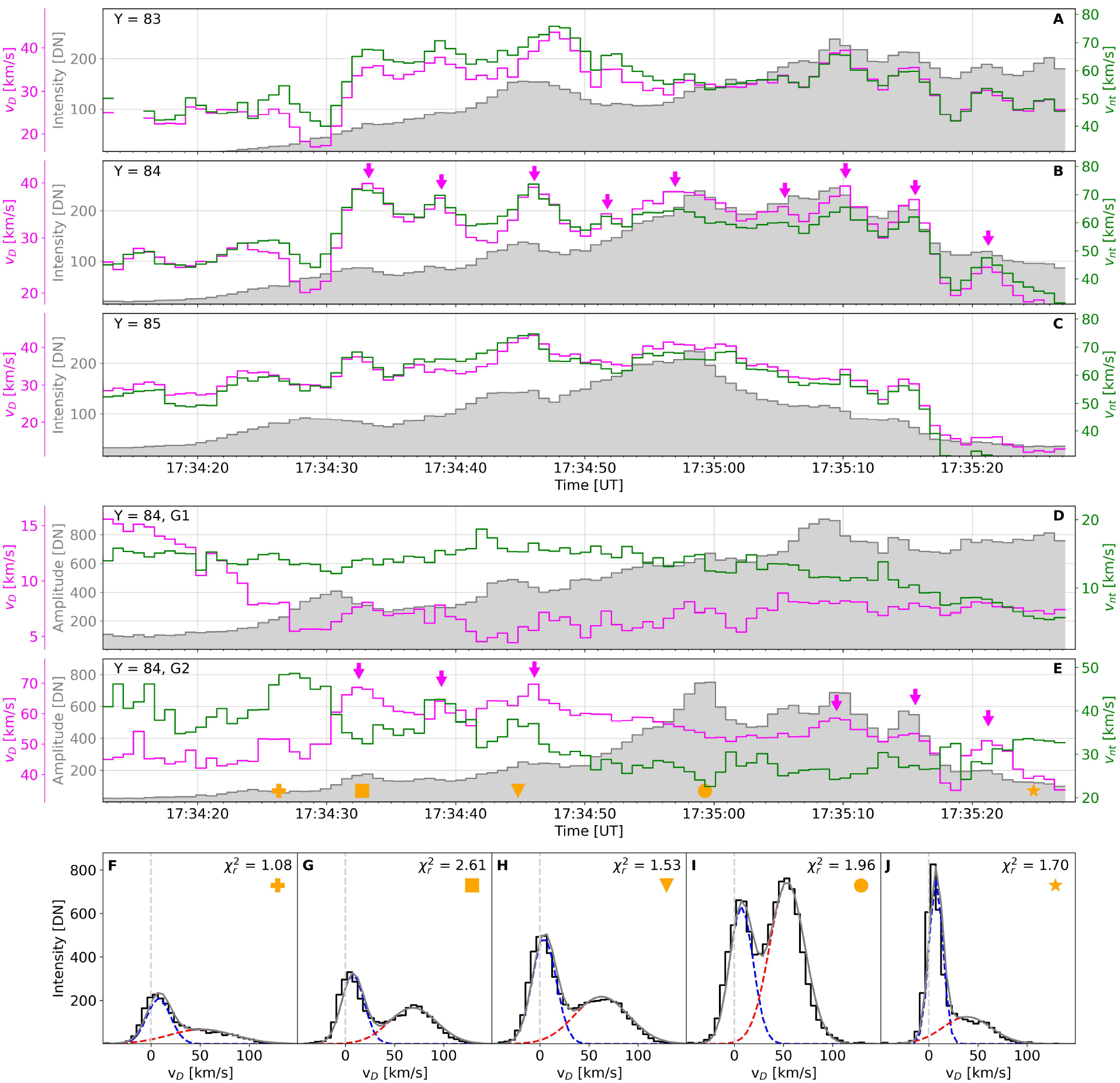}
\caption{Panels (A) -- (E) detail the time evolution of the total intensity (resp. {intensity} amplitude, grey), the Doppler velocity (magenta), and the non-thermal broadening (green) of the Si IV 1402.77\,\AA~line. The curves plotted in panels (A) -- (C) were obtained at 3 different pixel positions along the slit, 83 through 85, via the moment analysis. The time evolution of these quantities {at the pixel position 84 (black arrow in Figure \ref{fig_mom_maps}(A) -- (C)) obtained via two-Gaussian fitting is detailed in panels (D) and (E).} The first Gaussian (G1) is shown in panel (D), while the second Gaussian is shown in panel (E). {Magenta arrows in panels (B) and (E) highlight selected enhancements in the Doppler velocity determined from the moment analysis and Gaussian fitting, respectively.} The orange symbols plotted in panel (E) mark the instants of observations of profiles displayed in panels (F) -- (J). There, the observed profiles are plotted in black, the Gaussian fits to the near-at-rest and secondary redshifted components are plotted in blue and red, respectively, and the total fit to the profiles are in grey. In top-right corner of each panel, reduced chi-squared statistic of the respective fit is listed). }\label{fig_microscopic}
\end{figure*}

A brief statistical analysis of the profiles, performed using the k-means clustering implemented in the \texttt{scikit-learn} library, revealed that approximately 35\% of the profiles observed in the ribbon between 17:34 and 17:38 UT consisted of a primary component close to the rest wavelength of the line and a secondary redshifted component. {The pixels} where these spectra were found are plotted in Figure \ref{fig_mom_maps}(D) atop of the intensity contours corresponding to 10, 70, and 140\,DN (light-grey, grey, and black, respectively). After clustering $\approx3 \times 10^4$ profiles into 30 representative groups we found 4 groups which contained a majority of these profiles. The {mean spectra} belonging to these groups {(Nr. 9, 12, 18, 29)} are displayed in the bottom row of Figure \ref{fig_mom_maps}. The presence of the redshifted component was typically indicated by a pronounced red wing of the line. {The strength of this component was varying in both time and space. Profiles with the strongest secondary component are found in the group Nr. 9 (red), while relatively-weaker secondary components are present in profiles in groups Nr. 29 (magenta) and 12 (green). The group Nr. 18 (blue) contains roughly 18\% of the observed two-component spectra. The principal characteristic of the profiles of this group is the high separation of the two components as well as the strength of the secondary component, in some pixels even dominating the primary one (see also Figure \ref{fig_microscopic}(G) -- (I)).} A comparison of Figure \ref{fig_mom_maps}(C) and (D) shows that {these profiles} occurred in the two regions characterized by increased $v_\text{nt}$.

\subsubsection{Gaussian fitting of line profiles} \label{sec_fitting}

In order to precisely determine the properties of the spectra, the profiles observed in the ribbon were fitted using two Gaussians; one for the primary (`G1') and the other for the secondary redshifted (`G2') component. The fitting was performed using an automatic fitting routine \texttt{iris\_auto\_fit} included in the SolarSoft package. The time evolution of the {intensity} amplitudes (grey), Doppler velocities ($v_{\text{D}}$, magenta) of the centroids of the Gaussians, and their non-thermal broadening ($v_\text{nt}$, green) at the pixel position 84 {(Solar $Y$ = 171.25",} black arrow in Figure \ref{fig_mom_maps}(A) -- (C)) are plotted in Figure \ref{fig_microscopic}(D), (E). The time interval detailed in this figure was selected to depict the {short-term} evolution of the parameters of the Gaussians and corresponds to panels (A) -- (C) of the same figure. The orange symbols in panel (E) mark instants of observations of spectra plotted in Figure \ref{fig_microscopic}(F) -- (J). {These were selected as a representative sample of the spectra we analyze in this manuscript. A visual inspection of these profiles confirms that these spectra are indeed double-peaked and do not have a central reversal \citep[c.f. Figure 4][]{zhou22}, a manifestation of opacity effects in Si IV.}

As is clear from panel (D), the redshift of the G1 (magenta) initially dropped from $v_{\text{D}} \approx$15\,km\,s$^{-1}$ to roughly 7\,km\,s$^{-1}$ and stayed at this latter value during the remaining of the analyzed period. Similarly, $v_\text{nt}$ (green) showed only negligible variations until $\approx$17:35 UT when it started to progressively decrease. On the other hand, the {intensity} amplitude (grey) exhibited an overall increase during this time interval, with a few enhancements peaking at approximately  {17:34:30, 17:34:45}, and 17:35:08 UT. The properties of the G2 (panel (E)) varied significantly more than those of the G1. Oscillations qualitatively similar to those of the {short-term} evolution of the moments of the full line can be clearly seen in the $v_{\text{D}}$ curve between 17:34:25 and 17:34:50 UT, eventually reappearing $\approx$20 seconds later {(magenta arrows in panel (E))}. During two major enhancements peaking at 17:34:33 and 17:34:46 UT, $v_{\text{D}}$ reached nearly 70\,km\,s$^{-1}$. This is interesting as redshifts of transition region lines in flare ribbons exceeding 50\,km\,s$^{-1}$ are generally rare, a few exceptions were presented e.g. by \citet[][]{tian15, zhang16, li17}. Note that panels (B) and (E) of Figure \ref{fig_microscopic} show that the maximal Doppler shifts resulting from the moment analysis and fitting of the redshifted component, respectively, differ by nearly 40\%. This means that, because of the relative positions of centroids of the two components, the moment analysis systematically decreases the measured values of $v_{\text{D}}$ \citep[see also][]{tian15, lorincik22}. We suggest this may be the reason why large redshifts (typically derived from moments or single-Gaussian fits) of cool lines are only rarely reported. {Exceptionally high Si IV redshifts, in some cases exceeding 100\,km\,s$^{-1}$, have also been reported in IRIS observations of flare loops \citep[e.g.][]{tian14, zhou20, yu22}. These features are attributed to downward-oriented reconnection outflows. Since the ribbon we analyze here was not obscured by flare loops (Figure \ref{fig_loops_kernels}) it is very unlikely that the reconnection downflows could explain the observations we report on here.} $v_\text{nt}$ exhibited three clear enhancements between 17:34:25 and 17:34:50 UT, two of which had a counterpart in the $v_{\text{D}}$ curve. The highest number of short-term enhancements was exhibited by the {intensity} amplitude. This curve also exhibits {long-term} evolution consisting of an initial growth approximately until 17:35 UT, later followed by its decrease, similar to the total intensity for the full profile. 

On both {long- and short-term} scales, the parameters of the G2 follow trends comparable to those exhibited by the moments of the full spectral profile. Of particular interest is the striking similarity between the time evolution of the integrated intensity of the line (Figure \ref{fig_microscopic}(B)) and the amplitude of the G2 (Figure \ref{fig_microscopic}(E)). Similarly, the enhancements visible in $v_{\text{D}}$ of the G2 can easily be associated to those exhibited by the moments. On the other hand, only two enhancements of the non-thermal broadening of the G2 show a counterpart in $v_\text{nt}$ determined from the moment analysis (at 17:34:37 and 17:34:45 UT). 

\subsubsection{Correlation analysis}

\begin{figure}[h!]
\includegraphics[width=18cm]{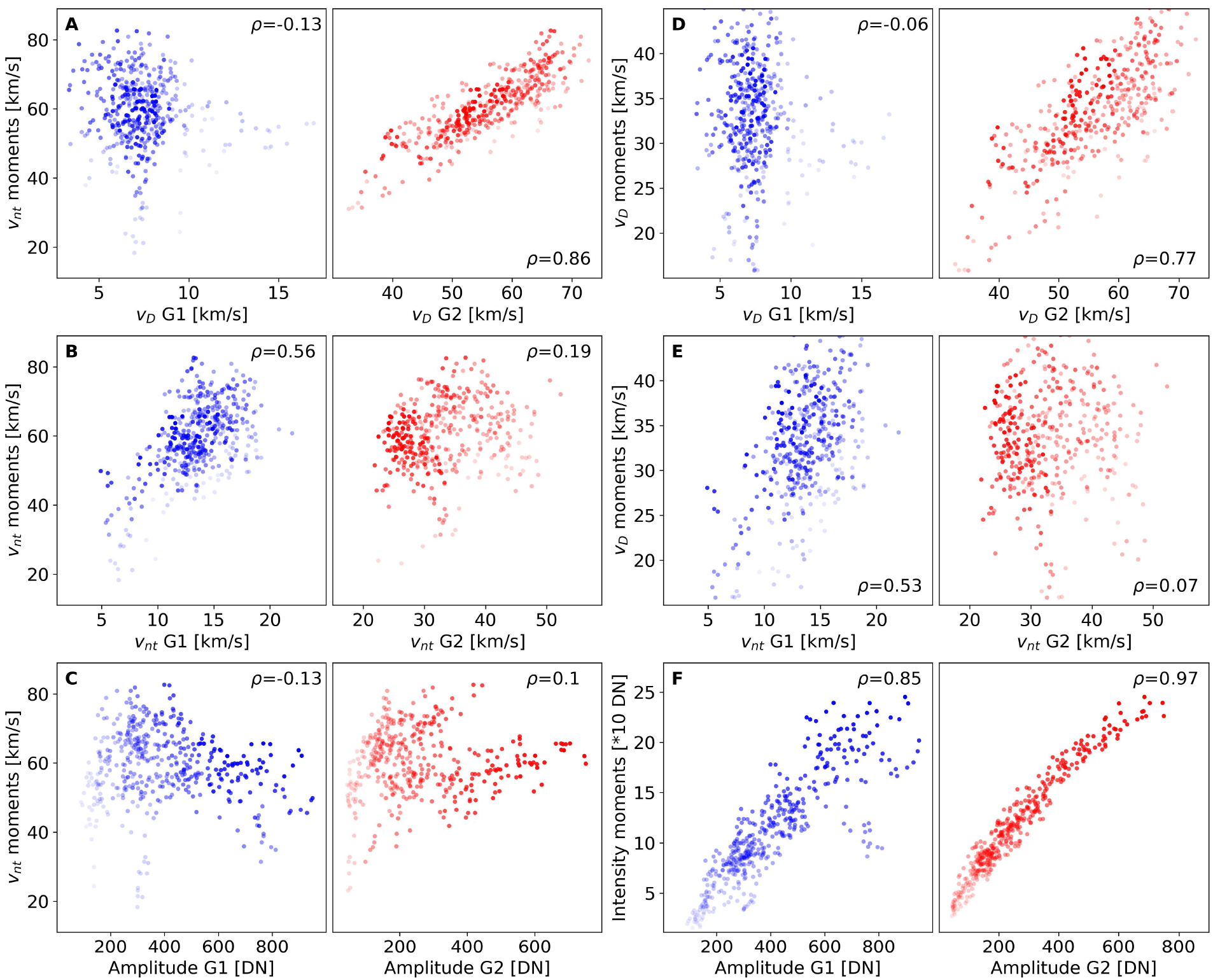}
\caption{Scatterplots detailing the relation between  Si IV 1402.77\,\AA~line properties determined via the moment analysis (vertical axes) and the parameters of the two Gaussians fitting its spectra (horizontal axes). In each panel, the Pearson correlation coefficient calculated in the respective scatterplot is listed.}\label{fig_correl}
\end{figure}

We analyzed further the dependence between the moments of the full Si IV 1402.77\,\AA~line and the double Gaussian fits using scatter plots. Pairs of scatterplots plotted in Figure \ref{fig_correl} detail relations between line properties determined via the moment analysis of the full line shown on the vertical axes and the three parameters of the G1 and G2 shown on the horizontal axes (blue and red columns, respectively). The data points (colored dots) plotted in these panels represent spectra corresponding to all of the cuts plotted in Figure \ref{fig_mom_maps}(A) -- (C), for those locations where the integrated intensity in a given pixel exceeded 10\,DN. The transparency of the dots depends on the total intensity of the profile, from the weakest spectra (transparent dots) to the strongest ones (opaque dots). In each panel, the Pearson correlation coefficient ($\rho$) for the respective scatter plot is shown. 

Panels (A) -- (C) of Figure \ref{fig_correl} detail the relations between the non-thermal broadening resulting from the moment analysis ($v_\text{nt, MA}$) and the three properties of the Gaussian fits. This column of panels indicates that $v_\text{nt, MA}$ exhibits the highest correlation with the Doppler shift of the G2 with $\rho = 0.86$ (panel (A)). $v_\text{nt, MA}$ shows a moderate correlation ($\rho = 0.56$) with $v_\text{nt}$ of the G1, whereas the correlation with the $v_\text{nt}$ of the G2 is weak, reaching only $\rho = 0.19$ (panel (B)). Only a very weak linear relation ($|\rho| \approx 0.1$) between the $v_\text{nt, MA}$ and the amplitudes of the Gaussians is indicated by the scatter plots in panel (C). These results imply that the enhancements seen in the $v_\text{nt, MA}$ curve are induced by varying redshifts of the line's secondary component rather than being due to a real broadening of the profile itself. This is in analogy with the results of \citet{depontieu10}, who proposed that quasi-periodic oscillations of intensity, velocity, and width of coronal lines, previously attributed to magnetoacoustic waves, could also be driven by quasi-periodic blueshifts induced by upflows at footpoints of coronal loops. 

Regarding other notable results from the correlation analysis, the $v_{\text{D}}$ values from the moment analysis show a relatively-high correlation with those of the G2 with $\rho = 0.77$ (panel (D)). On the other hand, we found no correlation ($\rho \approx 0$) between $v_{\text{D}}$ from the moment analysis and $v_{\text{D}}$ of the G1. According to panel (E), $v_{\text{D}}$ from the moment analysis is also moderately correlated with the non-thermal broadening of the G1 ($\rho = 0.53$). At the same time, no correlation with the broadening of the G2 was found. Finally, as seen in panel (F), the integrated intensity exhibits a slightly higher correlation with the amplitude of the G2 ($\rho = 0.97$) than with that of the G1 ($\rho = 0.85$).

\subsubsection{Observations and modeling of double-peaked Si IV line profiles} \label{sec_iris_discussion}

The results of the correlation analysis presented in the previous section indicate that the {short-term} evolution of the Si IV 1402.77\,\AA~line, which is characterized by quasi-periodic enhancements of its broadening as well as other properties, cannot be addressed by the presence or development of lower-atmospheric turbulences alone. We believe this is the case, despite the apparent similarity between the results of \citet{jeffrey18} and our observations of oscillatory signatures of the Si IV 1402.77\,\AA~line spectra. In \citet{jeffrey18}, the quasi-periodicity in broadening of apparently single-peaked profiles was interpreted in terms of the development of turbulence. However, in contrast to our results, the oscillatory trends reported by \citet{jeffrey18} were only observed in the broadening of the line. The authors concluded that time trends in which the bulk velocity does not exhibit oscillations can only be reproduced when multiple interacting Alfv\'en waves, representing velocity fluctuations, are present in the Si IV emitting region. We should also note that in their work individual or multiple components of the Si IV profiles were not discussed, so it is unclear whether those played a role. 

Profiles qualitatively similar to those we discuss here can be found in the analysis of \citet{brannon15}. In multiple instants detailed in their Section 3.3 and Figure 7, two components of the Si 1402.77\,\AA~exhibited a separation comparable to that we report on in Section \ref{sec_fitting}. In \citet{brannon15}, the component located closer to the rest wavelength showed weak Doppler shifts, in certain instants even blueshifts, which exhibit quasi-periodic variations similar to those of the secondary redshifted component {analyzed therein}. These are interpreted via elliptical waves propagating along reconnected field lines leading to minor oscillations of flare loops along the LOS. The discussed mechanisms expected to produce such waves are the Kelvin-Helmholtz or the tearing mode instability in the flare current sheet {\citep[see also][]{tian16}}. However, during the time period analyzed in the current paper (Section \ref{sec_fitting} and detailed Figure \ref{fig_microscopic}), the Doppler velocity of the G1 did not exhibit {short-term} evolution relatable to that of the G2. This indicates that the two components of the Si IV line were not affected by a common physical mechanism such as, for example, the instability-induced oscillations of flare loops. We hypothesize that the high {short-term} variability (timescales $< 10$\,s) of the parameters of the G2, dominating the G1 in terms of affecting the moments of the Si IV 1402.77\,\AA~line, suggests that the redshifted component mimicked the transient nature of Si IV spectra known from other analyses of flare ribbon emission. On the other hand, the negligible variations of $v_{\text{D}}$ and $v_\text{nt}$ of the G1 indicate that the primary component was less affected by the energy transport between the reconnection site and the lower atmosphere.  

Double-peaked profiles of IRIS lines are consistent with recent numerical simulations (calculated with the RADYN code) of the response of a flare atmosphere to heating by accelerated (non-thermal) electrons. \citet{kowalski17} showed that a model atmosphere that is heated using a high non-thermal energy flux of $F = 5 \times 10^{11}$\,erg\,cm$^{-2}$\,s$^{-1}$ contains two emitting layers in the chromosphere, one existing due to the ongoing condensation and the second, stationary, underneath it. {Even though the spectra synthesized using parameters of this atmosphere were double-peaked, we cannot immediately apply the results of this modeling to our observations. This is because, first, the authors did not provide predictions for the transition region emission and second, energy fluxes of this order of magnitude are more typical for stronger (X-class) flares \citep[see e.g.][]{kennedy15, kleint16, graham20}. We plan to focus on reproducing double-peaked Si IV spectra using models with electron beam parameters constrained by HXR observations in the future.} Still, these results illustrate that under specific conditions spectra of cool lines observed by IRIS might originate in two regions with distinct physical environments. Provided the Si IV line is formed under optically thin conditions, this could address the observed properties of the two-component Si IV line profiles. Recent extensions to the runs of \citet{kowalski17} showed that both the energy flux and the duration of the heating significantly affect the velocity profile of the chromospheric condensation in time, including its maximal velocity. As indicated in Figure 2 of \citet{kowalski22}, in one particular model these velocities reach up to more than $100$\,km\,s$^{-1}$ at the maximum gas mass density. To our knowledge, no observational support for condensation speeds of this order of magnitude has yet been presented. Whether the redshifts of the Gaussian fitting the secondary component of the line observed up to 70\,km\,s$^{-1}$ can be related to the condensation speeds resulting from this numerical simulation is discussed in Section \ref{sec_loops3D}. According to the models, the condensing plasma cools very rapidly and the condensation speeds of this order of magnitude last only for a brief period of time ($< 1$\,s), subsequently dropping to $\approx$25\,km\,s$^{-1}$ during the following $\approx$9\,s. The observed {short-term} evolution of $v_\text{D}$ of the G2 {(Figure \ref{fig_microscopic}(E))} consists of enhancements with a typical span between 3 to 7 time bins (2.4 to 5.6\,s), indicating the downflow durations of the same order of magnitude as predicted by the model. We finally note that apart from the large redshifts of the secondary component, the signatures of the response of the lower solar atmosphere to non-thermal particles were likely manifested in the increased $v_\text{nt}$. During the period analyzed in Figure \ref{fig_microscopic}(D) and (E), $v_\text{nt}$ of the G2 was typically 2 -- 3 times as large as that of the G1, what well corresponds to the observations of \citet{brannon15}. 

\section{Effects of projection on observed condensation speeds} \label{sec_loops3D}

As stated in Section \ref{sec_fitting}, the Si IV 1402.77\,\AA~line does not typically exhibit Doppler velocities as large as those of the centroid of the Gaussian fit to the redshifted component in our observations. The occurrence of Si IV redshifts at speeds of up to $\approx 70$\,km\,s$^{-1}$ is particularly interesting because the flare we report on occurred fairly close to the solar limb roughly at $\mu = 0.58$. It is thus of interest to consider whether the viewing geometry and observed velocities along the line-of-sight could be compatible with the velocities of the condensation-induced field-aligned downflows recently predicted by \citet{kowalski22}. 

\begin{figure}[h!]
\includegraphics[width=85mm, clip, viewport=10 610 1400 2000]{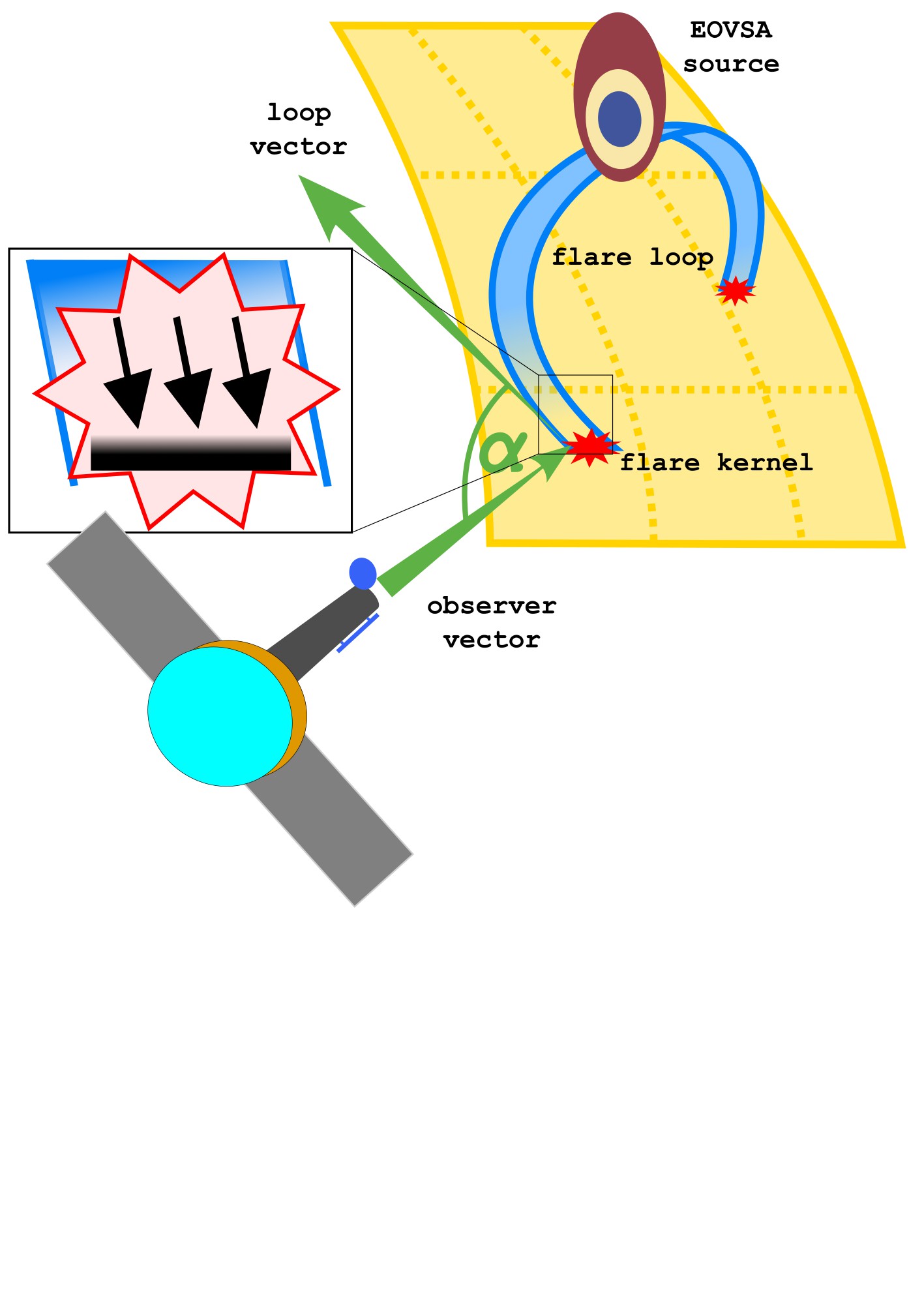}
\\
\caption{Cartoon demonstrating our definition of the angle $\alpha$ between the \textit{observer vector} pointing along the LOS of IRIS to the flare loop footpoint (flare kernel, red star) and the \textit{loop vector} directed from this footpoint to the nearest flare loop coordinate, i.e., a tangent to the loop footpoint. The zoomed inclusion on the left indicates two possible source regions of the observed spectra, one corresponding to the condensation layer (black arrows) and the other to the underlying stationary chromosphere (contoured bar). The maroon, yellow, and dark-blue contours indicate the source of the microwve emission measured by EOVSA.}\label{fig_cartoon}
\end{figure}

We address this issue by determining the viewing angles $\alpha$ of the flare loops within the flare loop arcade which are rooted at or close to the kernels which crossed the slit of IRIS during the analyzed period (Figure \ref{fig_loops_kernels}). We define $\alpha$ as the angle between two vectors illustrated in Figure \ref{fig_cartoon}. The first \textit{observer vector} points along the LOS of IRIS to the center of the disk. Since the angular distance between the kernel and the center of the disk is negligible compared to the distance between IRIS and the Sun, for simplicity we assume the vector to be pointing to the kernel itself. The second \textit{loop vector} points from the kernel toward the closest loop coordinate above the flare loop footpoint. Essentially, the loop vector is a tangent to the lowest segment of the reconstructed loop. The calculation of the viewing angle $\alpha$ was done in the Heliocentric Earth Equatorial (HEEQ) coordinate system after a conversion of the heliographic loop coordinates using formula (2) in \citet{thompson06}.

\begin{figure*}[h!]
\centering
\includegraphics[width=18cm, clip, viewport=20 20 1300 370]{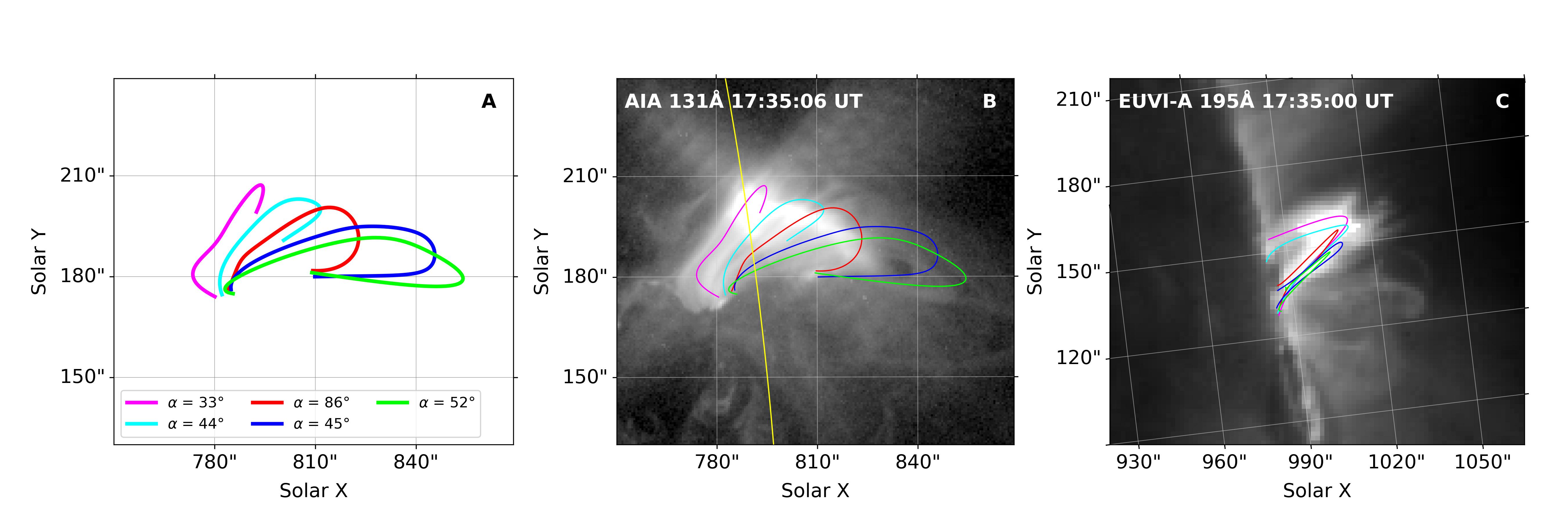}
\caption{Flare loops traced using stereoscopic observations of the flare arcade from the AIA 131\,\AA~and EUVI 195\,\AA~during the impulsive phase of the flare. Panel (A) shows these loops plotted in a blank image and lists the viewing angles $\alpha$ of these loops. Panels (B) and (C) show the loops replotted atop of snapshots from AIA and EUVI. The yellow line plotted in panel (B) indicates the location of the solar limb from the observing point of the STEREO-A spacecraft. }\label{fig_loops3D}
\end{figure*}

The loop reconstruction was performed using the \texttt{ssc\_measure} routine included in SolarSoft. This method is used to trace loops from imaging observations of two instruments, returning the Stonyhurst heliographic longitude, latitude, as well as the radial distance of the traced loop coordinates \citep[see e.g.][]{nistico13}. To do so we used observations of the flare loop arcade carried out in the AIA 131\,\AA~and EUVI 195\,\AA~filters during the impulsive phase of the flare. Even though the EUVI 195\,\AA~filter is primarily sensitive to plasma radiating at coronal temperatures (log $T$ [K] = 6.2), the evolution and the morphology of the arcade of flare loops was similar to that observed in the AIA 131\,\AA~(also compare Figure \ref{fig_overview}(F) and (G)). This is due to the temperature response of this channel, which, similarly as the AIA 193\,\AA~channel, contains a secondary peak at log $T$ [K] $\approx 7.2$ \citep[see Figure 11 in][]{wuelser04}. During our observation, the angular separation between SDO and STEREO-A was roughly 35$^\circ$. As seen in panel Figure \ref{fig_loops3D}(C), in this projection, EUVI observed the arcade at the solar limb which we also indicated in panel (B) using the yellow line. Most of the flare loops were oriented along the LOS of EUVI, which only let us use this instrument to constrain the height of the loops and their inclination from the vertical. Extracting the overall geometry of the individual loops among the arcade was thus largely contingent on the AIA 131\,\AA~image. The AIA 131\,\AA~data used for the tracing were therefore first averaged over 10 consecutive frames to suppress the noise and then sharpened using the Multi-Scale Gaussian Normalization \citep[MGN;][]{morgan14}. Still, portions of flare loops above their footpoints in the northern ribbon were obscured by the arcade. Even though the tracing of these loop segments was rather uncertain, this did not pose a limitation to our study as their viewing angles $\alpha$ are, for the most part, affected by the curvature and the orientation of the loops above the southern ribbon. Each of the loops we were able to trace was defined by 5 -- 10 manually-selected loop coordinates which we consequently interpolated using a third-order spline interpolation. Lastly, the conversion of the heliographic loop coordinates to the coordinate frames of AIA and EUVI, necessary for the visualisation of the loops, was performed using functions for transformations of coordinate frames implemented in {sunpy} \citep{sunpy20}. 

Five flare loops traced and reconstructed in 3D using the method described above are shown in Figure \ref{fig_loops3D}. The loops are replotted on top of a blank image (panel (A)) as well as AIA 131\,\AA~(panel (B)) and EUVI 195\,\AA~(panel (C)) snapshots acquired at the same time as those used for tracing of the loops. In the legend plotted in panel (A), the viewing angles $\alpha$ of the traced loops are listed. The smallest viewing angle of 33$^\circ$ corresponds to the magenta loop (leftmost): its portion just above the southern footpoint seems to be tilted towards the solar east direction. The largest viewing angle is that of the red flare loop, for which the lowest segment was nearly perpendicular to the LOS of IRIS with $\alpha = 86^\circ$. The viewing angles of the cyan, blue, and lime loops range between $\alpha = 44 - 52^\circ$. Using the estimated viewing angles of the flare loops rooted in the kernel, we can now estimate the field-aligned flows along the loop legs by deprojecting the Doppler velocities obtained from the centroids of the Gaussian fits to the secondary redshifted component of the line. The field-aligned flow velocity corrected for the viewing angle of the loop can simply be obtained as $v_\text{D,corr.} =  v_{\text{D}}/cos(\alpha)$. During the period detailed in Figure \ref{fig_microscopic}(E) the Doppler velocity of the G2 $v_{\text{D}}$ ranged between 30 and 70\,km\,s$^{-1}$. By using the average viewing angle of $\alpha = 52^\circ$ we obtain $v_\text{D,corr.}$ of 49 -- 114\,km\,s$^{-1}$. The upper limit of this range is close to the maximal condensation speeds recently predicted in RADYN simulations \citep[see Figure 2 in][]{kowalski22}. We note that the average value of $\alpha$ was significantly affected by the red flare loop with its large viewing angle that is far above the threshold of roughly $48^\circ$ that would lead to a value of $v_\text{D,corr.}$ that is within the constraints of the models. 

To wrap up the analysis of the viewing angles of the flare loops, even though our 3D reconstruction of flare loops was limited by several factors such as the orientation of the flare loop arcade along the LOS of EUVI, we were still able to trace several flare loops and estimate the corresponding field-aligned flows along the legs of these loops. These were found to be possibly consistent with those predicted in RADYN simulations.

\section{Microwave and SXR emission during the 2022 January 18 flare} \label{sec_radio}

\subsection{Sources of radio emission} \label{sec_sources}

\begin{figure*}[h!]
\centering
\includegraphics[width=18cm]{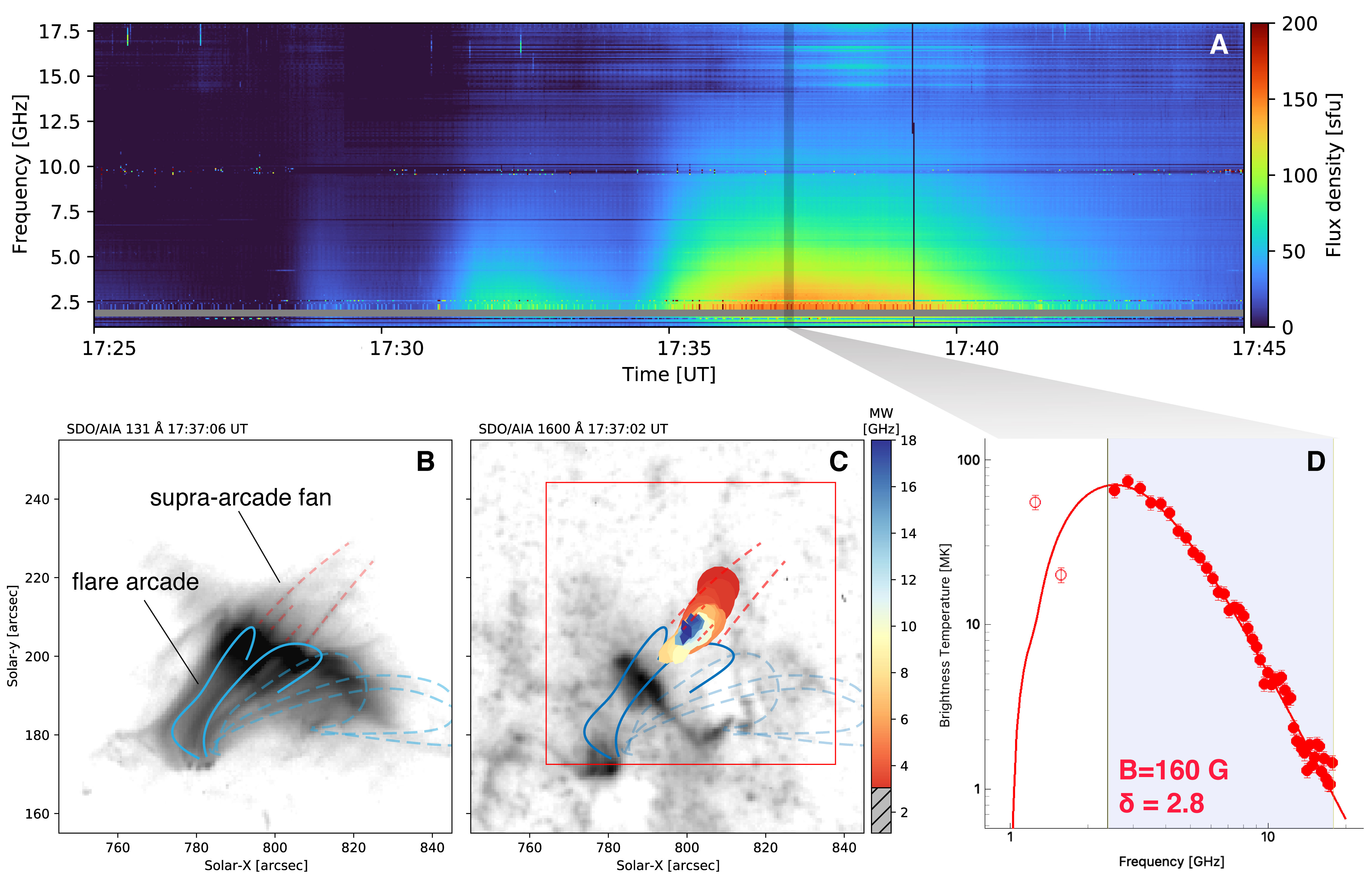}
\caption{EOVSA dynamic spectroscopic imaging of the looptop microwave source of the 2022 January 18 flare. Panel (A) shows EOVSA total (sun-as-a-star) power microwave dynamic spectrum of the flare in 1 -- 18\,GHz. In panel (B), reconstructed flare loops also shown in Figure \ref{fig_loops3D} are overplotted atop of AIA 131\,\AA~image. The red dashed curves denote the supra-arcade fan structure. The same flare loops and fan structure are indicated in panel (C) with the AIA 1600\,\AA~snapshot in the background. Between the apices of the two solid flare loops, sources of EOVSA microwave emission at 17:37 UT are plotted as contours corresponding to the 95\%~of the peak brightness at each frequency (decreasing from blue to red). Panel (D) presents the microwave brightness temperature spectrum $T_{B}(\nu)$ (red circles) obtained in the region denoted using the red box in panel (C). Each $T_{B}(\nu)$ value at a given frequency $\nu$ represents the maximum brightness temperature within the region. The red solid curve is the best-fit model based on non-thermal gyrosychrontron emission. The vertical white shaded area indicates the frequency range excluded from the spectral fit. The best-fit magnetic field strength $B$ and the power-law index $\delta$ of the non-thermal electron energy distribution are also shown.\label{fig_EOVSAspec}}
\end{figure*}

The EOVSA sun-as-a-star microwave dynamic spectrum after removing the pre-flare background (Figure \ref{fig_EOVSAspec}(A)) reveals several broadband impulsive bursts detected during this flare. The brightest microwave burst occurred at 17:37 UT and its flux density was over 150 sfu (solar flux units). The timing of the burst is denoted by the shaded vertical stripe in Figure \ref{fig_EOVSAspec}(A). The EOVSA images of the burst at frequencies between 3 and 18\,GHz are plotted in Figure \ref{fig_EOVSAspec}(C) using filled 95\% contours overlaid on the AIA 1600\,\AA~snapshot from the corresponding time. The contours were color-coded in radio frequency $\nu$ from red to blue with increasing frequency. The EOVSA sources at different frequencies appear to be distributed along a line with their higher-frequency end (blue colors) oriented toward the flare arcade. 

The highest-frequency source was located at the top of the flare arcade, enclosed by the two leftmost flare loops described in Section \ref{sec_loops3D} and plotted in Figure \ref{fig_loops3D}. These loops connect the northern ribbon with the brightest kernels in the southern ribbon, where the Si IV line observed by IRIS exhibited features consistent with the reconnection-induced chromospheric condensation (Section \ref{sec_iris_discussion}). The location and orientation of the source in height was also consistent with the supra-arcade fan structure seen in the 131\,\AA~image (red dashed lines in Figure \ref{fig_EOVSAspec}(B) and (C)). The latter has been commonly regarded as an observational indication of magnetic reconnection occurring above the flare arcade \citep{mckenzie99}. The location of the microwave source can be interpreted using the standard solar flare model \citep{masuda94}, in which the microwave emission is produced by the flare-accelerated electrons trapped in the newly-reconnected magnetic field lines that later form flare loops. The accelerated electrons propagate downward along the flare loops and, upon colliding with the chromosphere, lead to the development of downflows (Section \ref{sec_intro}) with signatures such as those we report on here. The dispersion of the EOVSA sources above the flare arcade can be attributed to the decrease of the coronal magnetic strength with height
\citep[for further discussions on EOVSA microwave spectroscopy see][etc.]{gary18,chen20cs}

The non-thermal nature of the microwave source was further affirmed by microwave spectral analysis. To investigate the nature of the microwave source, we derived the microwave brightness temperature spectrum $T_{B}(\nu)$ of the microwave source at 17:37 UT. In Figure \ref{fig_EOVSAspec}(D), we show the microwave spectra (red circles) derived from the red box in Figure \ref{fig_EOVSAspec}(C) at 17:37 UT. Each $T_{B}(\nu)$ value in the spectrum represents the brightness temperature maximum within the region at a given frequency $\nu$. The spatially-resolved microwave spectrum exhibited characteristics of incoherent non-thermal gyrosynchrotron emission produced by non-thermal electrons gyrating in the coronal magnetic field (compare our Figure \ref{fig_EOVSAspec}(D) and e.g. Figure 4(e) in \citet{chen20fr}, see also \citet{bastian98}). We have therefore adopted the gyrosynchrotron (GS) forward fit method described in \citet{fleishman20} to fit the brightness temperature spectrum $T_{B}(\nu)$ using an isotropic non-thermal electron source with a power-law energy distribution. We restricted our spectral fit to frequencies above 2\,GHz only, as the spectrum below 2\,GHz contains contributions from coherent radiation \citep{bastian98}. From the spectral fit, we obtained the magnetic field strength of 160 G and the non-thermal power-law index of 2.8. We note that the uncertainties of the best-fit parameters are not well constrained without a more in-depth evaluation, e.g., Monte Carlo analysis, which was beyond the scope of this study. Nevertheless, the microwave spectrum above 2\,GHz favors a non-thermal electron source presumably associated with particle acceleration-driven energy release in the flare.

\subsection{Comparison of SXR, radio, and UV lightcurves}

During the time period under examination, time derivatives of the SXR and microwave radio flux exhibited several enhancements which can be related to the {long-term} trends exhibited by Si IV spectra observed by IRIS. 

\begin{figure*}[h!]
\centering
\includegraphics[width=18cm]{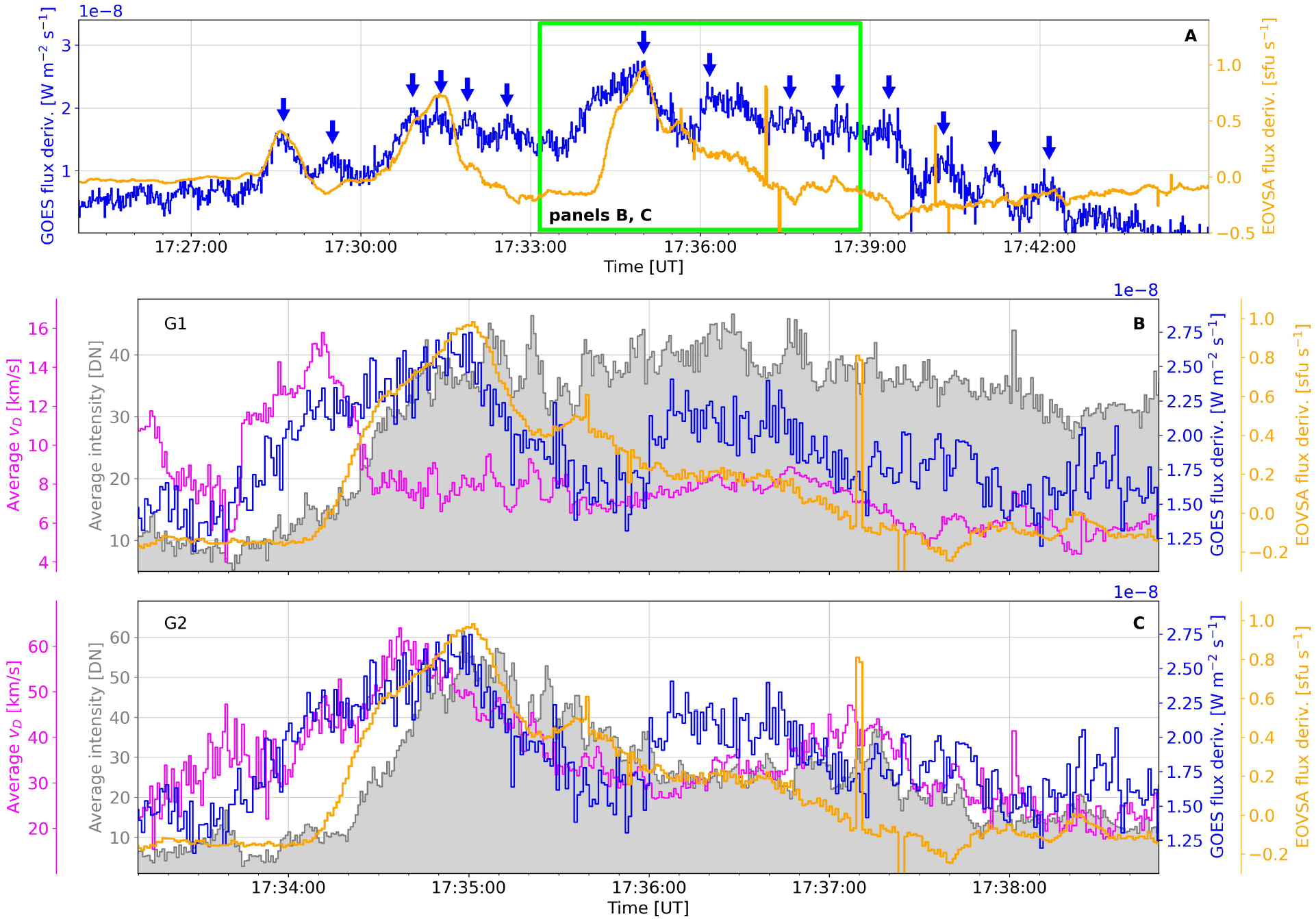}
\caption{Panel (A) demonstrates quasi-periodic pulsations (QPPs, {blue arrows}) observed during the impulsive and peak phases of the flare. The blue curve is the the time derivative of the GOES SXR flux in its 1 -- 8\,\AA~channel, while the orange curve corresponds to the microwave radio emission averaged in the range of 2.4 -- 5\,GHz observed by EOVSA. Panels (B) and (C) compare a section of these lightcurves (lime box in panel (A)) with properties of fits of two Gaussians fitting spectra of the Si IV 1402.77\,\AA~line. The grey filled curve represents time variations of the observed intensity, while the magenta curve corresponds to the Doppler velocity $v_\text{D}$ averaged in the ribbon. The GOES and EOVSA data plotted in this figure were produced using data smoothed with a 15\,s boxcar. }\label{fig_macroscopic}
\end{figure*}

Figure \ref{fig_macroscopic}(A) shows the time derivative of the SXR flux observed in the 1 -- 8\,\AA~channel of the GOES-16 satellite (blue) and the sun-as-a-star microwave emission measured by EOVSA averaged over frequencies between 2.4 and 5.0\,GHz (orange). To suppress the noise (also see Section \ref{sec_qpps}), both time derivatives were produced using data smoothed over 15\,s with a moving boxcar. In the period between 17:28 and {17:42} UT, the GOES time derivative exhibits numerous enhancements {(blue arrows)}, the most prominent one starting roughly after 17:33 UT, peaking at 17:35 UT, and decreasing until 17:36 UT. This peak, as well as peaks present at 17:28:30 and 17:31 UT, correspond to at least three enhancements visible in the EOVSA time derivative (cf. blue and orange curves). In panels (B) and (C) of the same figure, these time derivatives are compared with the properties of fits of the two components of the Si IV line. These panels detail the period between $\approx$17:33 -- 17:39 UT when the Si IV emission was enhanced (Section \ref{sec_mom_maps}) which increased the reliability of the fits to its components. The properties of the two Gaussians were averaged along the slit pixels crossing the ribbon, in this period corresponding to Solar $Y$ = 166" -- 176".

According to Figure \ref{fig_macroscopic}(B), the Doppler velocity (magenta curve) exhibited a sharp increase approximately between 17:33:40 -- 17:34:30 UT, peaking at $v_{\text{D}} \approx 16$\,km\,s$^{-1}$. This redshift later decreased, returning to its pre-increase values ranging between 5 -- 10\,km\,s$^{-1}$. The average intensity {(amplitude} $\times$ width, grey curve) of the G1 started to increase after 17:33:40 UT, peaked between $\approx$17:35 and 17:37 UT, and later slowly dropped. The increase of the average intensity corresponds to that of the EOVSA time derivative and reaches the maximum simultaneously with both EOVSA and GOES lightcurves (c.f. grey, orange, and blue curves). After peaking, both EOVSA and GOES lightcurves exhibit a decrease that is not visible in the properties of the G1. The enhancement of the intensity of the G2 was relatively shorter (Figure \ref{fig_macroscopic}(C)). After its onset after 17:34:20 UT and peak at 17:35 UT, the average intensity dropped to $\approx$50\% of its peak value during the following minute, remained roughly constant until 17:37:20 UT at which point it started to decrease. The peak of $v_\text{D}$ of this component occurred roughly 20 seconds prior to the peak of the intensity. In a trend similar to that of the average intensity, $v_\text{D}$ exhibited a decrease until 17:36 UT when it slightly rose again with a minor peak at 17:37:10 UT, and finally dropped. These curves show several interesting similarities with the EOVSA and GOES lightcurves. By starting to rise at 17:34:10 UT, peaking at 17:35 UT, and progressively dropping till 17:37:40 UT, the trend of the EOVSA lightcurve well matches that of the average intensity of the G2. The initial increase of the GOES time derivative observed before 17:35 UT resembles that of $v_{\text{D}}$. On the other hand, the peak of the GOES lightcurve corresponds to the peak exhibited by the average intensity at 17:35 UT. 

\subsection{Quasi-periodic pulsations} \label{sec_qpps}

Quasi-periodic enhancements are observed during flares across many passbands, typically at SXR, HXR, and microwave wavelengths. They are usually termed quasi-periodic pulsations \citep[QPPs; see e.g. the review of][and references therein]{nakariakov09}. QPPs are observed over a wide range of periods, from milliseconds to several minutes. From Figure \ref{fig_macroscopic}(A) alone it was rather hard to estimate the period of QPPs in this flare. The QPPs visible in the GOES time derivative in Figure \ref{fig_macroscopic}(A) exhibited four major and roughly ten minor peaks between 17:28 -- 17:42 UT, from which we estimate their period to 1 -- 3 minutes.

Numerous physical mechanisms have been associated with the generation of the QPPs \citep[see e.g. the review of][]{zimovets21}. \citet{mclaughlin18} divide these mechanisms into oscillatory and self-oscillatory processes. The oscillatory category includes for example natural MHD oscillations in flare loops and loops surrounding them or driving of periodic QPPs by external MHD waves. The self-oscillatory processes concern for example periodic or repetitive magnetic reconnection. The duration and periods of QPPs driven by the self-oscillatory processes are given by the timescales of the energy release during the reconnection. As summarized by \citet{hayes19}, current observations do not provide conclusive answers on the origin of the driver of QPPs, noting that various {processes} possibly act in different flares and their phases. 

We hypothesize that the QPPs we report on, notably the prominent one peaking approximately at 17:35 UT, were most likely driven by magnetic reconnection. Our study provides twofold evidence for this scenario, in both cases dealing with signatures of non-thermal electrons in multi-wavelength observations presented in Sections \ref{sec_spectra} -- \ref{sec_radio}. First, the SXR time derivative of GOES data reaches a maximum simultaneously with the time derivative of the EOVSA microwave emission which features are consistent with a non-thermal electron source (Section \ref{sec_sources}). This observable has also been reported by \citet{yusijie20} who claimed that the correspondence between microwave and X-ray bursts supports a scenario in which the 5-minute period QPPs observed therein were entirely driven by the reconnection. Second, the presence of the accelerated particles is indicated by the double-peaked Si IV 1402.77\,\AA~spectra observed by IRIS. The peak of the intensity trend of the redshifted component, most-likely existing as a consequence of the chromospheric condensation (Section \ref{sec_iris_discussion}), corresponds to those of the EOVSA and GOES time derivatives. It ought to be noted that similar conclusions were reached by \citet{li15qpp} who reported on simultaneous observations of 4-minute period QPPs and peaks in redshifts as well as broadening of C I, O IV, Si IV, and Fe XXI lines observed by IRIS. We also speculate that if these QPPs were generated by MHD waves or oscillations in or near the reconnection site, the two components of the Si IV line would exhibit similar time evolution. For example, the quasi-periodic enhancements would likely be visible in the Doppler shifts of both components \citep[as in][]{brannon15} which we did not observe. 

Note that apart from the {long-term} evolution of the two Gaussians, namely of the G2 fitting the redshifted component, we also observed oscillatory signals in the spectral properties of the Si IV line on timescales below 10\,s (Section \ref{sec_spectra}). Given the apparent correlation of the low-frequency oscillations of Si IV with the GOES and EOVSA signals, it is tempting to speculate whether such a correlation also explains the signals with shorter periods. The GOES time derivative did indeed exhibit omnipresent high-frequency oscillations, best visible in Figure \ref{fig_macroscopic}(B) and (C). However, the amplitudes of these oscillations were usually reaching a few $10^{-9}$\,W\,m$^{-2}$\,s$^{-1}$, while $\sigma$ of this curve in the same time interval was around $\approx 1\times10^{-9}$\,W\,m$^{-2}$\,s$^{-1}$. We could thus not distinguish whether these oscillations were induced by the flare or were simply signatures of the noise \citep[see][for SXR irradiance noise analysis using data from older GOES satellites]{simoes15}. Since IRIS captures only a few kernels at one time under the slit, and the flare extends over a larger FOV than IRIS can capture, it is quite possible that the full-disk GOES signal mixes information from several sources at any time, thus leading to "solar" noise that would not be expected to be correlated with the spectral properties IRIS observed in the few kernels. The relations between high-frequency oscillations in Si IV spectra and SXR flux time derivatives deserve attention of future studies, especially those analyzing flares with short impulsive phases \citep[see e.g.][]{hudson21} where fast rise time of GOES SXR flux is translated into relatively-higher values of the time derivatives (although such observations would still be limited by the full-disk nature of the signal). 

\section{Summary and conclusions}  \label{sec_summary}

In this manuscript we presented observations of the 2022 January 18 M-class flare, the first major flare the IRIS satellite observed at a sub-second cadence. We primarily analyzed trends exhibited by the intensity, Doppler velocity $v_\text{D}$, and non-thermal broadening $v_\text{nt}$ of the transition region Si IV 1402.77\,\AA~line in one of the ribbons during the impulsive phase of the flare. 
We paid particular attention to {short-term} evolution (scales of seconds) that these quantities, determined via the moment analysis, showed when the slit crossed the ribbon. The intensity of the line exhibited high-frequency oscillations with an estimated period of 10\,s. The period of oscillations seen in $v_\text{D}$ and $v_\text{nt}$ was roughly 7\,s. A substantial portion of profiles observed in the analyzed ribbon yielded double-peaked spectra. The primary component of the line observed close to its rest wavelength exhibited modest redshifts below $v_\text{D}$ = 15\,km\,s$^{-1}$. The redshifts of the secondary component were higher, temporarily reaching \mbox{$v_\text{D}$ $\approx70$\,km\,s$^{-1}$}. For a brief period of time the two components showed a high separation ($> 60$\,km\,s$^{-1}$), making the components well-resolved. The time evolution of the total intensity as well as $v_\text{D}$ and $v_\text{nt}$ resulting from the moment analysis was correlated with parameters of the Gaussian fit to the redshifted component of the line. Of particular interest was a strong correlation we found between $v_\text{nt}$ resulting from the moment analysis and $v_\text{D}$ of the redshifted Gaussian. This suggests that the signatures of the oscillatory enhancements in the broadening of the line were driven by quasi-periodicities in the downward-oriented motions of plasma visible in the secondary component of the line. This result, combined with other properties of the Si IV spectra observed in this event (see below), suggests that the oscillations of line broadening do not appear to be caused by plasma turbulence in the same way as in \citet{jeffrey18} (Section \ref{sec_iris_discussion}). At the same time, the properties of the primary component did not exhibit oscillatory trends comparable to those of the secondary component. Because of the inherently different behavior of the two components we proposed that the quasi-periodic patterns were not induced by oscillations of flare loops driven by instabilities in the reconnection region, an interpretation used by \citet{brannon15} for another flare that also exhibited rather rare double-peaked Si IV spectra.

Apart from the large redshifts, the likely origin of the secondary component was indicated by the broadening of the Si IV line. The fact that $v_\text{nt}$ of this component exceeded that of the primary component by a factor of 2 -- 3 suggested that the redshifted component existed due to an ongoing energization by magnetic reconnection, compatible with \citet[][]{brannon15}. Based on the different behaviour of both redshifts and broadening of the two components we speculate that the Si IV emission possibly originated from two regions at the flare footpoints with different physical conditions, in line with RADYN simulations of \citet{kowalski17}. Our results appear to be compatible with a scenario in which the relatively-broader redshifted component could be linked to the condensation region and the relatively-narrower primary component to the underlying stationary region. Confirming this scenario would however require a proper simulation of the transition region emission. Nevertheless, the highest redshifts of the secondary component are consistent with the maximal velocities of chromospheric condensation ($\gtrapprox 100$\,km\,s$^{-1}$) recently found in RADYN simulations \citep{kowalski22}. The estimated field-aligned condensation speeds were found by estimating viewing angles of flare loops anchored in the analyzed kernels based on a simple flare loop reconstruction in 3D. 

Finally, we found that {long-term} trends (scales of minutes) of the intensity and Doppler shift exhibited by the secondary component were well matched by one of the QPPs observed in time derivatives of SXR (GOES) and microwave radio (EOVSA) fluxes. This result hints that the driving mechanism of the QPPs observed during this flare was repeated magnetic reconnection that also led to the formation of the secondary component of the Si IV line through the chromospheric condensation. Both EOVSA and GOES time derivatives also showed high-frequency oscillations. The question whether these enhancements were manifestations of high-frequency QPPs relatable to the oscillations of Si IV line properties remains to be answered in the future.

\section*{Conflict of Interest Statement}

The authors declare that the research was conducted in the absence of any commercial or financial relationships that could be construed as a potential conflict of interest.

\section*{Author Contributions}

The research project was led by J.L. He was primarily engaged in the processing and visualization of data from IRIS, EOVSA, and GOES, as well as the preparation of the manuscript. V.P. and B.D.P. oversaw the progress of work on the project and primarily contributed to the interpretation of the observations. B.D.P. designed the very high-cadence IRIS observing programs and proposed the detailed study of these observations. S.Y. was involved in the processing and analysis of radio data from EOVSA. N.F. was providing support necessary for processing and visualization of data in Python. All co-authors reviewed this manuscript and agreed with its submission. 

\section*{Funding}

This work was supported by NASA under contract NNG09FA40C ({\it IRIS}). V.P. also acknowledges support from NASA grant 80NSSC20K0716. S.Y. was supported by NASA grant 80NSSC20K1283 to NJIT. N.F. was also supported by NASA contract NNG04EA00C (\textit{SDO}/AIA).

\section*{Acknowledgments}

The authors are grateful to Hugh Hudson and Paulo Sim\~oes for their insights on analysis of SXR data, Markus Aschwanden for tips on reconstruction of loops in three dimensions, Adam Kowalski for discussions regarding the chromospheric condensation, and Alberto Sainz Dalda for assistance with the k-means clustering techniques. IRIS is a NASA small explorer mission developed and operated by LMSAL with mission operations executed at NASA Ames Research center and major contributions to downlink communications funded by ESA and the Norwegian Space Centre. AIA and HMI data are provided courtesy of NASA/SDO and the AIA science team. 

\onecolumn

\bibliographystyle{Frontiers-Harvard} 
\bibliography{bibliography}

\clearpage

\section*{Figures}

\end{document}